# Distributed Control with Low-Rank Coordination

Daria Madjidian and Leonid Mirkin

*Abstract*—A common approach to distributed control design is to impose sparsity constraints on the controller structure. Such constraints, however, may greatly complicate the control design procedure. This paper puts forward an alternative structure, which is not sparse yet might nevertheless be well suited for distributed control purposes. The structure appears as the optimal solution to a class of coordination problems arising in multi-agent applications. The controller comprises a diagonal (decentralized) part, complemented by a rank-one coordination term. Although this term relies on information about all subsystems, its implementation only requires a simple averaging operation.

## I. INTRODUCTION

The ability to cope with complexity is one of the challenges of control engineering nowadays. Already an established research area by the late 1970s [1]–[3], control of complex systems reinvigorated during the last decade, impelled by recent technological progress, networking and integration trends, efficiency demands, etc.

Complexity may be manifested through different attributes, one of which is the presence of a very large number of sensors and actuators. In such situations fully centralized, structureless, information processing becomes infeasible. This motivates the quest for distributed control methods, with various constraints on information exchange between subsystems and information processing in the controller. Such structural constraints are conventionally expressed in terms of sparsity pattern [3]–[5], with nonzero elements corresponding to permitted coordination between subsystems. Sometimes, delay constraints on the communication between subsystems are considered [5].

Although indeed natural, sparsity (and / or delay) constraints substantially complicate the analysis and, especially, design of control systems. Many well understood problems might turn acutely opaque when sparsity constraints on the controller are added [6], [7]. Analysis is simplified if the plant happens to possess a compatible sparsity pattern (the quadratic invariance condition [5], [8]) or if additional constraints are imposed on the closed-loop behavior (like positivity [9], [10]). But even then the computational burden grows rapidly with the problem dimension and, more importantly, structural properties of the resulting controller are rarely transparent. Revealing such properties proved to be a challenge even in relatively simple problems, see [11], [12] and the references therein.

This research was supported by the Swedish Research Council through the LCCC Linnaeus Center and by the European commission through the project AEOLUS.

D. Madjidian is with the Department of Automatic Control, Lund University, Box 118, SE–221 00 Lund, Sweden. E-mail: *daria@control.lth.se*.

L. Mirkin is with the Faculty of Mechanical Eng., Technion—IIT, Haifa 32000, Israel. E-mail: *mirkin@technion.ac.il*.

This paper puts forward an alternative structure. We study a class of large-scale coordination problems that happens to admit a solution of a different type: not sparse, but nevertheless potentially feasible in various distributed control applications. Specifically, we consider a homogeneous group of autonomous agents, i.e., a group of systems having identical dynamics and identical local criteria. Coordination requirements are then introduced through a (global) linear constraint imposed on an "average" agent. This setting is motivated by certain control tasks arising in the control of wind farms. If no sparsity constraints are imposed, the problem admits an analytic solution endowed with two appealing properties. First, the computational burden in this setting is independent of the number of agents. Second, the optimal feedback gain is of the form of a block-diagonal matrix perturbed by a block-rank-one component. The structures of these components are transparent. The diagonal part merely comprises the local, uncoordinated, gains. The rank-one part is then responsible for coordination via fine-tuning the local controllers on the basis of measurements of an "average" agent.

The (weighted) averaging is the only non-sparse, centralized, task that has to be performed by the controller. We argue that this task may be network-friendly too. The averaging is a relatively simple numerical operation, which might be robust to sensor imperfections for large groups. It can be performed either locally, by each agent, or globally, by a coordinator. The averaging of measured variables of individual subsystems may be viewed as a spatial counterpart of the generalized sampling operation [13]. This is in contrast to the decentralized structure, which may be thought of as a form of the ideal sampling, which ignores the intersample information. Considering this analogy, it might even be useful to impose the control structure in problems where, unlike in our formulation, it does not appear as a property of the optimal solution. In fact, one such approach, also in the context of large-scale systems, was proposed in [14], see Remark 3.5 for more details.

The paper is organized as follows. In Section II we consider a coordination problem arising in wind farms. This problem serves a motivation for the theoretical developments in Sections III (problems with hard coordination constraints) and IV (soft constraint formulations). Both sections illustrate their developments by numerical studies of the same wind farm coordination problem. Section III also contains an extensive discussion on properties of the resulting controller configuration and the structure of the optimal cost. Concluding remarks are then provided in Section V.

*Notation:* The transpose of a matrix $M$ is denoted as $M'$. By $e_i$ we understand the $i$th standard basis of an Euclidean space and by $I_n$—the $n \times n$ identity matrix (we drop the dimension subscript when the context is clear). The notation $\otimes$ stands for the Kronecker product of matrices, see [15, Ch. 13]. The $L^2(\mathbb{R}^+)$ norm [16, Ch. 4] of a signal $\xi$ is denoted as $\|\xi\|_2$.

## II. MOTIVATING EXAMPLE: COORDINATION IN WIND FARMS

Wind energy is an increasingly active application area for control, see [17] and the references therein. Lately, the focus is shifting from control of a stand-alone wind turbine (WT) to coordinated control of networks of WTs, commonly known as wind power plants (WPP) or wind farms. In this section we consider a coordination problem arising in large-scale WPPs, which is used to motivate the problem studied in this paper.

### A. Problem description

We consider the problem discussed in [18], [19], where a WPP is required to meet a certain power demand. To achieve this, the WTs need to coordinate their power production. Since there are multiple WTs in the farm, certain freedom exists in distributing the power demand among them. This freedom can be used to address local objectives of individual turbines, such as regulating rotor speed, reducing fatigue loads, preventing excessive pitch action, etc. Thus, instead of following a fixed portion of the power demand, a WT can be allowed to continuously adjust its power production in response to local wind speed fluctuations. Since wind speed fluctuations are not the same across the WPP, changes in power production that benefit one WT can be compensated for by changes at WTs with opposite needs.

For control design purposes, it is common practice to model a WT as a linear system around an operating point. It may also be natural to make two additional simplifying assumptions.

1) WTs in a WPP are often identical in their design. By assuming that they operate around the same mean wind speed and mean power production, the WTs may be considered to have equal dynamics.
2) Due to a large distance between individual turbines in WPPs, it may be assumed that wind speed variations experienced by them are uncorrelated [20], [21].

With these observations in mind, below we address a coordination problem among a group of $\nu$ WTs. For simplicity, we use a stripped-down[1] version of the individual WT model and performance index studied in [21]. The model is derived from [22] and describes an NREL 5-MW wind turbine [23], operating around a mean wind speed of 10 m/s and a nominal power production of 2 MW. Each WT is assumed to be equipped with an internal controller, which manipulates the blade pitch angle and generator torque in order to track an external power reference. At the nominal power production, the WT operates in the derated mode (below maximum power production) and is able to both increase and decrease its power production. The turbine models are given by

$$\dot{x}_i = Ax_i + B_w w_i + B_u u_i, \quad i = 1, \ldots, \nu$$

[1]We measure the input in MW and use neither the dynamic model of the effective wind speed (its DC gain is absorbed into the model) nor dynamic weights on regulated signals (we use approximate static weights instead).

where $\begin{bmatrix} A & B_w & B_u \end{bmatrix}$ take the following numerical values:

$$\begin{bmatrix} 0 & 120 & -0.92 & 0 & 0 & 0 & 0 \\ 0.0084 & -0.032 & 0 & 0 & 0 & 0.12 & -0.021 \\ 0 & 150 & -1.6 & 0 & 0 & 0 & 0 \\ 0 & 0 & 0 & 0 & 1 & 0 & 0 \\ 0.021 & 0.054 & 0 & -4 & -0.32 & 0.2 & 0 \end{bmatrix}.$$

Here the state vector spells out as

$$x_i = \begin{bmatrix} \text{pitch angle} \\ \text{rotor speed} \\ \text{internal controller state} \\ \text{nacelle fore-aft position} \\ \text{nacelle fore-aft speed} \end{bmatrix}$$

and the exogenous disturbance $w_i$ is the deviation in wind speed from its nominal value, modeled as a white noise process with unit intensity. The control signal $u_i$ is the deviation in the power reference from its nominal value. The model neglects generator dynamics, which makes $u_i$ equal to the actual deviation in the power production of the WT.

Following [21], we assume that each turbine aims at achieving a trade-off between regulating the rotor speed, reducing fatigue loads on the tower, and preventing excessive pitch activity and power deviations. The performance of the $i$th turbine is quantified as the variance of the regulated variable $z_i = C_z x_i + D_{zu} u_i$, where

$$\begin{bmatrix} C_z & D_{zu} \end{bmatrix} = \begin{bmatrix} \mathrm{diag}\{\sqrt{0.1}, 100, 0, 100, 0\} & 0 \\ 0 & 1 \end{bmatrix}.$$

In other words, for each turbine we consider the state-feedback $H^2$ problem for the closed-loop system from $w_i$ to $z_i$.

The combined power production of the WTs must satisfy a power demand to the WPP, which is assumed to be the sum of nominal WT power productions. Since $u_i$ is the deviation from nominal WT power production, this requirement can be imposed as the constraint

$$\sum_{i=1}^{\nu} u_i = 0, \tag{1}$$

which introduces coordination between individual WTs.

The resulting constrained $H^2$ problem can be converted to a standard unconstrained one by resolving (1) for any $i$, say as $u_1 = -(u_2 + \cdots + u_\nu)$. This results in an $H^2$ problem with $\nu$ subsystems and $\nu - 1$ control signals. Yet the dynamics of subsystems and the cost function in this problem are coupled. This might, especially if the number of turbines in the WPP is very large, considerably complicate both the solution procedure (the curse of dimensionality) and the implementation of the resulting controllers. Therefore a *scalable* solution procedure is of interest.

### B. Towards a scalable solution

As discussed in the Introduction, the conventional approach in the field is to impose some kind of sparsity constraints on the controller and seek a scalable optimization procedure to solve it. By limiting the information exchange between subsystems, a sparse structure can ensure that the information processing at each subsystem remains viable as the number of subsystems grows. This property is important, so it frequently preponderates over inevitable losses of performance. The problem is that imposing sparsity constraints might significantly complicate the design. Once the constraint (1) is resolved, our problem only satisfies the quadratic invariance condition of [8] for a handful of structural constraint options (e.g., block triangular). Another choice discussed in the Introduction, imposing positivity constraints on the closed-loop dynamics [9], is not engineeringly justified for our problem because we work in deviations from nominal values. We thus may consider resorting to non-convex optimization procedures, relying upon a proper choice of initial parameter guess.

To provide a flavor of such an approach, we confine our attention to static state-feedback controllers, $u = Fx$, and add the constraint $F \in \mathcal{H}_\eta$, where for a given $\eta \in \mathbb{N}$

$$\mathcal{H}_\eta := \{F : F_{ij} = 0 \text{ whenever } |i - j| > \eta\}$$

and the addition in the spatial variable is performed modulo-$\nu$ (e.g., $\nu + 1 = 1$). We then use an approach, similar to that proposed in [19], which, in turn, makes use of the distributed gradient method of [24].

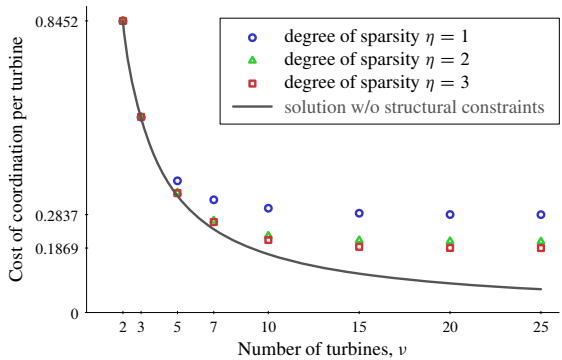

Fig. 1. Cost of coordination per turbine under different degrees of sparsity.

Fig. 1 shows the normalized difference between the $H^2$ performances attained with and without the coordination constraint (1) (the normalized cost of coordination) as a function of $\nu$ for different degrees of sparsity $\eta$. We can see that as the sparsity restriction is relaxed, i.e., as $\eta$ increases, the performance improves. Besides, the performance of sparse controllers improves as $\nu$ increases. We can also see that the improvement is not as fast as in the non-sparse solution (shown by the solid line). This, however, may be well expected and is not the main focus of this example.

Rather, we would like to emphasize difficulties encountered in designing the sparse controllers. Although not visible on the plot, these difficulties are readily appreciable. The fact that the problem is not convex ($\mathcal{H}_\eta$ is not quadratically invariant under this plant) renders the whole procedure sensitive to the choice of initial values for the feedback gain. We did experience convergence to local minima, so the solutions presented in Fig. 1 are the result of multiple runs of the algorithm. And we still cannot guarantee that the results are globally optimal[2]. In addition, the optimization procedure itself is quite demanding, its computational complexity grows with the increase of $\nu$. Finally, the results of the numerical procedure are not transparent, with no indication of what effect small changes of the system parameters might have on it.

To conclude, from the distributed control viewpoint the problem appears to be a challenge. Nonetheless, in the next section we show that it has a closed-form solution, which is computationally scalable and possesses additional appealing properties from the distributed control perspective.

### III. LQR WITH COORDINATION CONSTRAINTS

Motivated by the problem considered in Section II, in this section we study an optimization problem for non-interacting subsystems, having identical dynamics, with coordination constraints. To simplify the presentation, we consider an LQR version of the problem, although the extension to the $H^2$ formalism (external disturbances) is straightforward.

#### A. Problem statement

Consider $\nu$ independent systems

$$\Sigma_i : \dot{x}_i(t) = Ax_i(t) + Bu_i(t), \quad x_i(0) = x_{i0} \quad (2)$$

where $x_i(t) \in \mathbb{R}^n$ can be measured, $u_i(t) \in \mathbb{R}^m$, and $A \in \mathbb{R}^{n \times n}$ and $B \in \mathbb{R}^{n \times m}$ are such that the pair $(A, B)$ is stabilizable. Associate with each of these systems the performance index

$$\mathcal{J}_i = \int_0^\infty \bigl(x_i'(t) Q_\alpha x_i(t) + u_i'(t) u_i(t)\bigr) \mathrm{d}t \quad (3)$$

for some $n \times n$ matrix $Q_\alpha \geq 0$ such that the pair $(Q_\alpha, A)$ has no unobservable modes on the $j\omega$-axis. Minimizing $\mathcal{J}_i$ for $\Sigma_i$ would be a set of $\nu$ standard uncoupled LQR problems. We couple these problems by constraining the behavior of the *center of mass* of $\Sigma_i$, understood as the system

$$\bar{\Sigma} : \dot{\bar{x}}(t) = A\bar{x}(t) + B\bar{u}(t), \quad \bar{x}(0) = \bar{x}_0 \quad (4)$$

connecting the signals

$$\bar{u}(t) := \sum_{i=1}^\nu \mu_i u_i(t) \quad \text{and} \quad \bar{x}(t) := \sum_{i=1}^\nu \mu_i x_i(t), \quad (5)$$

where the weights $\mu_i \neq 0$ may be thought of as the masses of each subsystem. Coordination is then imposed by requiring $\bar{\Sigma}$ to evolve according to

$$\dot{\bar{x}}(t) = (A + B\bar{F})\bar{x}(t), \quad \bar{x}(0) = \bar{x}_0 \quad (6)$$

for a given gain $\bar{F} \in \mathbb{R}^{m \times n}$. This yields the following problem formulation:

$$\text{minimize} \quad \mathcal{J} := \sum_{i=1}^\nu \mathcal{J}_i \quad (7a)$$

$$\text{subject to} \quad \Sigma_i, \quad i = 1, \dots, \nu \quad (7b)$$

$$\bar{u} - \bar{F}\bar{x} = 0 \quad (7c)$$

where $\bar{u}$ from (7c) substituted into (4) yields (6). In addressing (7), we implicitly restrict our attention to stabilizing controllers only. Without loss of generality, we also assume that the weights are normalized as $\sum_i \mu_i^2 = 1$.

---

[2]In fact, they are not, as attested by the sub-optimality of the resulting cost in the case of $\eta = 3$ and $\nu = 7$, for instance.

*Remark 3.1 (connections with the motivating problem):* It is readily seen that the problem considered in Section II is a particular case of (7) corresponding to $\bar{F} = 0$. Constraint (7c) can thus also be viewed as a constraint imposed on average trajectories. ▽

*Remark 3.2 (minimizing weighted sum of $\mathcal{J}_i$):* The weights $\mu_i$ may be manipulated to assign importance to each subsystem. This can also be attained via replacing $\mathcal{J}$ in (7a) with the weighted sum $\mathcal{J} = \sum_i \lambda_i \mathcal{J}_i$ for some $\lambda_i > 0$. The addition of $\lambda_i$, however, does not enrich the design. It is only a matter of scaling each $x_i$ and $u_i$ by $\sqrt{\lambda_i}$ and then replacing $\mu_i$ with $\mu_i/\sqrt{\lambda_i}$ (with the normalization assumption $\sum_i \mu_i^2/\lambda_i = 1$). In the choice between '$\mu$' and '$\lambda$' scalings we picked the former because it allows negative weights. ▽

### B. Problem solution

We start with rewriting (7) in an aggregate form using the Kronecker product notation. Introduce the unit vector

$$\mu := \begin{bmatrix} \mu_1 & \cdots & \mu_\nu \end{bmatrix}'$$

and the aggregate state and control signals $x := \sum_i e_i \otimes x_i$ and $u := \sum_i e_i \otimes u_i$, respectively. In this notation, the dynamics of the aggregate systems are

$$\dot{x}(t) = (I_\nu \otimes A)x(t) + (I_\nu \otimes B)u(t), \quad (8)$$

the cost function in (7a) is

$$\mathcal{J} = \int_0^\infty \bigl(x'(t)(I_\nu \otimes Q_\alpha)x(t) + u'(t)u(t)\bigr)\mathrm{d}t, \quad (9)$$

and the constraint (7c) reads

$$(\mu' \otimes I_m)u - (\mu' \otimes \bar{F})x = 0, \quad (10)$$

The key idea behind our solution is to apply coordinate transformations to the state and input signals that decouple constraint (7c) on the one hand, while preserving the uncoupled structure of the system and cost on the other. This can be achieved by the coordinate transformations

$$\tilde{x} := (U \otimes I_n)x \quad \text{and} \quad \tilde{u} := (U \otimes I_m)u \quad (11)$$

for some *unitary* matrix $U \in \mathbb{R}^{\nu \times \nu}$. Indeed, using the relation $(M_1 \otimes N_1)(M_2 \otimes N_2) = (M_1 M_2) \otimes (N_1 N_2)$, it is readily seen that both (8) and (9) remain the same, modulo the replacement of $x$ and $u$ with $\tilde{x}$ and $\tilde{u}$, respectively, while the coordination constraint changes and becomes

$$(\mu'U' \otimes I_m)\tilde{u} - (\mu'U' \otimes \bar{F})\tilde{x} = 0.$$

To achieve decoupling, we may consider the following requirements on $U$:

$$U\mu = e_1. \quad (12)$$

Because $\mu$ is assumed to be a unit vector, there is always a $U$ satisfying (12). A possible choice is the matrix of transpose left singular vectors of $\mu$.

Thus, when expressed in terms of $\tilde{x}$ and $\tilde{u}$ with $U$ satisfying (12), problem (7) still has an uncoupled cost function and uncoupled dynamics. But now the constraint, which reads $(e_1' \otimes I_m)\tilde{u} - (e_1' \otimes \bar{F})\tilde{x} = 0$, is imposed only upon the first elements of $\tilde{u}$ and $\tilde{x}$, i.e., it reduces to

$$\tilde{u}_1 - \bar{F}\tilde{x}_1 = 0. \quad (13)$$

Hence, (7) splits into $\nu$ independent problems, with the $i$th problem depending only on the variables $\tilde{x}_i$ and $\tilde{u}_i$.

For $i = 2, \ldots, \nu$, we have identical unconstrained LQR problems with dynamics of the form (2) and cost functions of the form (3). Each one of these problems is then solved by the (stabilizing) control laws $\tilde{u}_i(t) = F_\alpha \tilde{x}_i(t)$, where $F_\alpha := -B'X_\alpha$ and $X_\alpha \geq 0$ is the stabilizing solution of the algebraic Riccati equation (ARE)

$$A'X_\alpha + X_\alpha A + Q_\alpha - X_\alpha BB'X_\alpha = 0. \quad (14)$$

These control laws achieve the optimal performance $\tilde{x}_{i0}' X_\alpha \tilde{x}_{i0}$.

For $i = 1$, constraint (13) completely determines $\tilde{u}_1$, as $\tilde{u}_1 = \bar{F}\tilde{x}_1$, rendering the optimization irrelevant. The plant dynamics then become

$$\dot{\tilde{x}}_1(t) = (A + B\bar{F})\tilde{x}_1(t), \quad \tilde{x}_1(0) = \tilde{x}_{10}$$

and the cost function

$$\int_0^\infty \tilde{x}_1'(t)(Q_\alpha + \bar{F}'\bar{F})\tilde{x}_1(t)\mathrm{d}t.$$

The dynamics of $\tilde{x}_1$ are stable iff $A + B\bar{F}$ is Hurwitz and in this case the value of the cost function is finite and equals $\tilde{x}_{10}' \bar{X} \tilde{x}_{10}$, where $\bar{X} \geq 0$ verifies the Lyapunov equation

$$(A + B\bar{F})'\bar{X} + \bar{X}(A + B\bar{F}) + Q_\alpha + \bar{F}'\bar{F} = 0. \quad (15)$$

The arguments above solve (7) in terms of the transformed variables in (11). What is left is to transform this solution back to $x$ and $u$. This is done in the following theorem, which is the main technical result of this section:

*Theorem 3.1:* Let $A + B\bar{F}$ be Hurwitz and the pair $(Q_\alpha, A)$ have no unobservable pure imaginary modes. Then the ARE (14) and the Lyapunov equation (15) are solvable, with $\bar{X} \geq X_\alpha$, and the unique solution of (7) is

$$u_i(t) = F_\alpha x_i(t) + \mu_i(\bar{F} - F_\alpha)\bar{x}(t), \quad (16)$$

where $F_\alpha = -B'X_\alpha$ is the LQR gain, associated with the uncoordinated version of the problem, without (7c), and $\bar{x}$ is the state vector of the center of mass $\bar{\Sigma}$ defined by (5). The optimal performance attainable by this controller is

$$\mathcal{J}_{\mathrm{opt}} = \sum_{i=1}^\nu \mathcal{J}_{i,\mathrm{opt}} + \bar{x}_0'(\bar{X} - X_\alpha)\bar{x}_0, \quad (17)$$

where $\mathcal{J}_{i,\mathrm{opt}} = x_{i0}' X_\alpha x_{i0}$ is the optimal uncoordinated costs of $\Sigma_i$ and $\bar{x}_0$ is the initial condition of the center of mass.

*Proof:* The solvability of the Riccati equations under the conditions of the theorem is a standard result [16, Thm. 13.7]. The inequality $\bar{X} \geq X_\alpha$ follows by the fact that if $u_i = \bar{F}x_i$, then $\mathcal{J}_i = x_{i0}' \bar{X} x_{i0} \geq x_{i0}' X_\alpha x_{i0} = \mathcal{J}_{i,\mathrm{opt}}$ for any $x_{i0}$. Now, the developments preceding the formulation of the theorem imply that the optimal control law for the transformed system is $\tilde{u} = \tilde{F}\tilde{x}$, where

$$\tilde{F} = (I_\nu - e_1 e_1') \otimes F_\alpha + (e_1 e_1') \otimes \bar{F}.$$

Then (11) implies that the optimal control law for the aggregate problem (8)–(10) is $u = Fx = (U' \otimes I_m)\tilde{F}(U \otimes I_n)x$, so, with the help of (12), we end up with the optimal gain

$$F = I_\nu \otimes F_\alpha + (\mu\mu') \otimes (\bar{F} - F_\alpha), \quad (18)$$

which yields (16). Finally,

$$\mathcal{J}_{\text{opt}} = \tilde{x}_0'((I_\nu - e_1 e_1') \otimes X_\alpha + (e_1 e_1' \otimes \bar{X}))\tilde{x}_0 \quad (19a)$$
$$= x_0'(I_\nu \otimes X_\alpha + (\mu\mu') \otimes (\bar{X} - X_\alpha))x_0, \quad (19b)$$

from which (17) follows immediately. □

*Remark 3.3 (constraining a part of $\bar{u}$):* If $\bar{F} = F_\alpha$, then the Lyapunov equation (15) is solved by $\bar{X} = X_\alpha$ and (16) reduces to the decentralized control law solving the uncoordinated version of (7). In other words, the coordination constraint becomes void if it attempts to mimic the optimal unconstrained dynamics. Likewise, we can constrain only a part of $\bar{u}$ by mimicking the optimal, with respect to (3), control trajectory of the partially constrained problem by its other part. Namely, let $E$ be a tall matrix such that $E'E = I$. It can be shown that the optimization of (7), with (7c) replaced by the partial constraint $E'\bar{u} - \bar{F}_1\bar{x} = 0$, corresponds to the original formulation with

$$\bar{F} = E\bar{F}_1 - (I - EE')B'X_2$$

where $X_2 \geq X_\alpha$ is the stabilizing solution of the ARE

$$(A + BE\bar{F}_1)'X_2 + X_2(A + BE\bar{F}_1)$$
$$+ (Q_\alpha + \bar{F}_1'\bar{F}_1) - X_2 B(I - EE')B'X_2 = 0$$

and the stabilizability of the pair $(A + BE\bar{F}_1, B(I - EE'))$ is required. Equation (15) is solved then by $\bar{X} = X_2$. ▽

### C. Discussion

The remainder of this section is devoted to properties of the solution presented in Theorem 3.1. In particular, we discuss the structure of the optimal controller and its suitability for distributed control applications (§III-C1), interpret the LQR problems in terms of the transformed variables (11) arising in the derivation as a technical step (§III-C2), quantify the effect of the coordination constraint (7c) on the performance of each subsystem (§III-C3), and explore the possibility of adding tracking requirements to the behavior of the center of mass (§III-C4).

*1) Control law: computation and structure:* An important property of the solution of Theorem 3.1 is its computational scalability. To calculate the optimal controller, we only need to solve ARE (14), which is the Riccati equation associated with the local, unconstrained, LQR. The *computational effort* to obtain the solution is thus independent of the number of subsystems $\nu$, which is an attractive property in the context of distributed control.

The low computational burden is not the only property of controller (16) that is appealing in distributed control applications. Its structure is even more intriguing. The optimal control law is a superposition of a local term, $F_\alpha x_i(t)$, and a (scaled) coordination term,

$$u_{\text{coord}}(t) := (\bar{F} - F_\alpha)\bar{x}(t). \quad (20)$$

The former is the optimal uncoordinated control law for $\Sigma_i$ and is fully decentralized. Coordination then adds a "correction" of the form $\mu_i u_{\text{coord}}$ to this local controller. This term destroys the (sparse) decentralized structure as none of the elements of the overall feedback gain (18) is zero in general. Nonetheless, the resulting configuration might suit large-scale applications well.

The non-sparse coordination term, which may be thought of as a (block) rank-one correction to the (block) diagonal local controller (cf. (18)), depends only on the behavior of the center of mass. Thus, although this term hinges upon information about all subsystems, the only operation required in its construction is averaging. This information clustering may be thought of as a form of *spatial generalized sampling* where the information required to form the correction component, $u_{\text{coord}}$, is obtained by aggregating distributed information in a weighted average.

The information aggregation via $\bar{x}$ is clearly less demanding, from both computation and communication viewpoints, than an individual processing of each $x_i$. Hence, the control law (16), although centralized, may be feasible for distributed control. Measurements of the center of mass could, in principle, be done either globally, by a coordinator, or even locally, by each subsystem.

*Remark 3.4 (an interpretation of the coordination policy):* Constraint (7c) can be satisfied without information exchange if each subsystem applies $u_i = \bar{F}x_i$. The term $(\bar{F} - F_\alpha)x_i$ can then be interpreted as a desired violation of this strategy in order to improve the performance with respect to $\mathcal{J}_i$. By rewriting the coordination term (20) as

$$u_{\text{coord}}(t) = \sum_{i=1}^{\nu} \mu_i (\bar{F} - F_\alpha) x_i(t),$$

we see that exchanging information (coordination) allows the subsystems to compensate for each other's violations. ▽

*Remark 3.5 (earlier appearance):* The diagonal-plus-low-rank configuration has already been used in [14], also in the context of control of large-scale systems. The motivation and technical tools used there, however, are quite different from those studied in this paper. The "low-rank centralized correction" to block-diagonal controllers is introduced in [14] to enlarge the design parameter space in the context of robust control of interconnected systems. The parameters are then designed via an LMI procedure, which utilizes some of the degrees of freedom brought about by this addition. In our setup, the structure *results from* an optimization problem and is responsible for coordinating otherwise uncoupled subsystems. As a result, our low-rank term is transparent, with clearly traceable effect on control performance (see below). ▽

*2) LQR problems in terms of $\tilde{x}_i$ and $\tilde{u}_i$:* The transformation of state and input coordinates defined by (11) and (12) serves the purpose of decomposing the problem into one problem with a prespecified control law and and $\nu - 1$ unconstrained LQRs. These problems have meaningful interpretations.

First, a comparison of (13) and (7c) suggests that

$$\tilde{x}_1 = \bar{x} \quad \text{and} \quad \tilde{u}_1 = \bar{u}.$$

This is indeed true, as can be seen through $\tilde{x}_1 = (e_1' \otimes I_n)\tilde{x} = ((e_1'T^{-\prime}) \otimes I_n)x = (\mu' \otimes I_n)x = \bar{x}$, for instance. Thus, the constrained problem is concerned with the center of mass (4) and its solution results in the dynamics as in (6), as expected.

The other components of $\tilde{x}$ and $\tilde{u}$ do not possess such interpretations per se, they are not even unique. Nevertheless, the unconstrained LQR cost built on them,

$$\tilde{\mathcal{J}} := \sum_{i=2}^{\nu} \int_0^\infty \bigl(\tilde{x}_i'(t)Q_\alpha \tilde{x}_i(t) + \tilde{u}_i'(t)\tilde{u}_i(t)\bigr)\mathrm{d}t$$

(this is what the control law (16) actually minimizes), can be interpreted. To this end, rewrite

$$\sum_{i=2}^{\nu} \tilde{u}_i'\tilde{u}_i = \tilde{u}'((I - e_1 e_1') \otimes I_m)\tilde{u} = u'((I_\nu - \mu\mu') \otimes I_m)u$$

(the last equality is obtained by (11) and (12)) and, likewise, $\sum_{i=2}^{\nu} \tilde{x}_i' Q_\alpha \tilde{x}_i = x'((I_\nu - \mu\mu') \otimes Q_\alpha)x$. It can be shown, by routine regrouping, that

$$I_\nu - \mu\mu' = \sum_{i=1}^{\nu} (e_i - \mu_i\mu)(e_i - \mu_i\mu)' \quad (21a)$$

$$= \sum_{i=1}^{\nu-1} \sum_{j=i+1}^{\nu} (\mu_j e_i - \mu_i e_j)(\mu_j e_i - \mu_i e_j)', \quad (21b)$$

Form (21a),

$$\tilde{\mathcal{J}} = \sum_{i=1}^{\nu} \int_0^\infty \bigl((x_i - \mu_i\bar{x})'Q_\alpha(x_i - \mu_i\bar{x}) + (u_i - \mu_i\bar{u})'(u_i - \mu_i\bar{u})\bigr)\mathrm{d}t.$$

In other words, $\tilde{\mathcal{J}}$ may be thought of as the cost of deviating from the normalized center of mass. The normalization becomes particularly transparent if all systems have equal masses, i.e., if $\mu_i = 1/\sqrt{\nu}$. In this case $\mu_i\bar{x} = \frac{1}{\nu}\sum_i x_i$ and $\mu_i\bar{u} = \frac{1}{\nu}\sum_i u_i$ are merely the average state and input signals and $\tilde{\mathcal{J}}$ quantifies the cumulative deviation from the average. In the same vein, (21b) leads to

$$\tilde{\mathcal{J}} = \sum_{i=1}^{\nu-1} \sum_{j=i+1}^{\nu} \int_0^\infty \bigl((\mu_j x_i - \mu_i x_j)'Q_\alpha(\mu_j x_i - \mu_i x_j) + (\mu_j u_i - \mu_i u_j)'(\mu_j u_i - \mu_i u_j)\bigr)\mathrm{d}t,$$

which penalizes mutual deviations of each subsystem from the others (the scaling factors $\mu_i$ and $\mu_j$ just align the subsystems to render the comparison meaningful), thus encouraging the achievement of an optimal *consensus*.

Summarizing, by solving (7) we effectively reach two goals: impose a required behavior on the center of mass and minimize discrepancy between subsystems. The optimal $\tilde{\mathcal{J}}$ can then be viewed as a measure of "gregariousness" or, perhaps, as a "herd instinct index" in the aggregate system (8). It follows from the proof of Theorem 3.1 (cf. (19a)) that

$$\tilde{\mathcal{J}}_{\mathrm{opt}} = \tilde{x}_0'\bigl((I_\nu - e_1 e_1') \otimes X_\alpha\bigr)\tilde{x}_0 = x_0'\bigl((I_\nu - \mu\mu') \otimes X_\alpha\bigr)x_0$$

$$= \sum_{i=1}^{\nu} \mathcal{J}_{i,\mathrm{opt}} - \bar{x}_0' X_\alpha \bar{x}_0. \quad (22)$$

Thus, the attainable local uncoordinated costs $\mathcal{J}_{i,\mathrm{opt}}$ also determine the cumulative closeness of systems $\Sigma_i$ to each other. It is worth emphasizing that $\tilde{\mathcal{J}}_{\mathrm{opt}}$ does not depend on the constraint imposed on the behavior of the center of mass. This separation is an intriguing property of the solution of (7).

*3) Cost of coordination per subsystem:* The last term in the right-hand side of (17) quantifies the deterioration of the (aggregate) performance $\mathcal{J}$ due to the coordination constraint (7c). Below, we look into the effect of coordination on the performance of individual subsystems.

We begin with the following result:

*Proposition 3.2:* The value of the $i$th performance index $\mathcal{J}_i$ under the control law (16) is

$$\mathcal{J}_i = \mathcal{J}_{i,\mathrm{opt}} + \mu_i^2 \bar{x}_0'(\bar{X} - X_\alpha)\bar{x}_0, \quad (23)$$

where $\bar{x}_0$ is the initial condition of the center of mass.

*Proof:* The control law (16) is a superposition of the locally optimal control law and the signal $v_i = \mu_i(\bar{F} - F_\alpha)\bar{x}$. It is known (see the proof of [16, Thm. 14.2]) that for any $v_i$,

$$\mathcal{J}_i = \mathcal{J}_{i,\mathrm{opt}} + \|v_i\|_2^2.$$

As follows from (6), the last term in the right-hand side above equals $\mu_i^2 \bar{x}_0' X_v \bar{x}_0$, where $X_v \geq 0$ solves the Lyapunov equation

$$(A + B\bar{F})'X_v + X_v(A + B\bar{F}) + (\bar{F} - F_\alpha)'(\bar{F} - F_\alpha) = 0.$$

(23) then follows by the fact that $X_v = \bar{X} - X_\alpha$, which can be verified by straightforward algebra. □

The second term in the right-hand side of (23) is exactly the cost of coordination for the $i$th subsystem. It is a function of the other subsystems through the vector $\bar{x}_0$. The dependence of $\bar{x}_0$ on an unspecified relation between the initial states of all subsystems complicates the analysis of the cost of coordination. If, for instance, $\bar{x}_0 = 0$, then $\mathcal{J}_i = \mathcal{J}_{i,\mathrm{opt}}$ and the coordination in that case comes at no cost. But if every $x_{i0} = \mu_i \chi$ for some $\chi \in \mathbb{R}^n$, then $\bar{x}_0 = \chi$ and we end up with $\mathcal{J}_i = x_{i0}' \bar{X} x_{i0}$. This is what we would have if the control laws $u_i = \bar{F} x_i$ were applied to each subsystem, which would correspond to an attempt to enforce (7c) without communication between subsystems. To avoid the dependence on $\bar{x}_0$, we assume through the rest of this subsection that $\bar{x}_0$ is bounded as a function of the number of subsystems $\nu$. In this case the term $\bar{x}_0'(\bar{X} - X_\alpha)\bar{x}_0$ is bounded as well and the cost of coordination becomes quadratically proportional to the corresponding "mass" $\mu_i$.

Consider now what happens with the cost of coordination per subsystem when the number of subsystems $\nu \to \infty$. It follows from the normalization assumption $\sum_{i=1}^{\nu} \mu_i^2 = 1$ that at most a finite number subsystems may have $\mu_i \not\to 0$ in this case. If such subsystems do exist, they dominate (5) and we then effectively have coordination between a finite number of subsystems. It is then natural that the cost of coordination for those subsystems does not vanish as $\nu$ grows. If, however, all $\mu_i \to 0$ as $\nu \to \infty$, the situation is different. In this case the coordination constraint (7c) is, in a sense, spread among all subsystems and the cost of coordination per subsystem vanishes with the increase of $\nu$. For example, if we assign equal weights to each subsystems, i.e., if every $\mu_i = 1/\sqrt{\nu}$, then the coordination toll per subsystem decreases inversely proportional to the number of subsystems. The decrease of the coordination cost is intuitive, as the addition of more subsystems brings more opportunities for coordination.

*4) Tracking:* Constraint (7c) can be modified to incorporate tracking requirements on the center of mass (4). For example, we may consider the constraint

$$\bar{u} = \bar{F}\bar{x} + r$$

for an exogenous signal $r$ (e.g., it may be a function of a reference signal). This would yield the control law

$$u_i(t) = F_\alpha x_i(t) + \mu_i(\bar{F} - F_\alpha)\bar{x}(t) + \mu_i r(t),$$

instead of (16) and the following dynamics of the center of mass:

$$\dot{\bar{x}}(t) = (A + B\bar{F})\bar{x}(t) + Br(t)$$

in lieu of (6). The cost function $\mathcal{J}$ in this case is no longer relevant per se, it might even be unbounded. Still, the "measure of gregariousness" interpretation of the unconstrained part of the optimization, as discussed in §III-C2, remains valid. Moreover, the value of the cost function in (22) is finite and independent of $r$, so it can be used to quantify group tracking properties of the system.

Fig. 2. Tadpoles tracking a point via rank-one coordination

An example of this group tracking capability is shown in Fig. 2. This animation presents a group of $\nu = 50$ "tadpole" agents, aiming at tracking a red object. Each agent is modeled as a second-order system, with uncoupled integrator dynamics in $x$ and $y$ directions, so that $A = 0_{2\times 2}$ and $B = I_2$. We choose all $\mu_i = 1/\sqrt{\nu}$, the local LQR weights to result in $F_\alpha = -I_2$, and pick $\bar{F} = -25 I_2$, so that the dynamics of the center of mass are considerably faster than those of each agent. The reference signal, showed as a red point in Fig. 2 and yielding the "$\infty$" shape in the $x$–$y$ plane, is

$$x_{\text{ref}}(t) = \begin{bmatrix} \sin(4\pi t) \\ 0.25 \sin(8\pi t) \end{bmatrix}.$$

To track this reference signal by each agent, we have chosen $r = \sqrt{\nu}\,[\bar{T}(0)]^{-1} x_{\text{ref}}$, where $\bar{T}(s) := (sI - A - B\bar{F})^{-1} B$. During the simulation, white noise disturbances were added to each agent to produce more vivid motion.

*D. Wind farm example (cont'd)*

We are now in the position to return to the example studied in Section II. To render the current LQR problem formulations compatible with that in §II-A, we assume that $x_{i0} = B_w v_i$, where $v_i$ are mutually independent random variables of unit variance. This yields $\bar{x}_0 = B_w \bar{v}$, where $\bar{v} := \sum_i \mu_i v_i$ is of unit variance as well. We then end up with (7) with $B = B_u$, $Q_\alpha = C_z' C_z$, $\bar{F} = 0$, and $\mu_i = 1/\sqrt{\nu}$ for all $i$.

By Theorem 3.1, the optimal control law is given by

$$u_i = F_\alpha x_i - F_\alpha x_a,$$

where $x_a := \frac{1}{\nu}\sum_i x_i$ is the average state of wind turbines and the gain $F_\alpha$ is obtained by solving ARE (14). To calculate the cost of coordination depicted in Fig. 1 by the solid line, we use Proposition 3.2 to end up with the formula

$$\mathcal{J}_i - \mathcal{J}_{i,\text{opt}} = \tfrac{1}{\nu} B_w'(\bar{X} - X_\alpha) B_w,$$

where $\bar{X}$ is the observability Gramian of $(C_z, A)$. This cost tends to zeros as $\nu \to \infty$.

With its structural properties revealed, the non-sparse solution to (7) compares favorably with the sparsity-based one considered in §II-B. Our calculations are scalable, in fact, they are independent of the number of turbines. The result is always globally optimal. The effect of the coordination constraint on the local performance of each turbine is transparent and easy to calculate as well. The price we pay is that the resulting controller is centralized. This might not be feasible in some situations where communication constraints are restrictive. Still, the only centralized information processing that is required to execute the control law is the averaging operation to calculate $F_\alpha x_a$. This does not require an individual processing of the global state of the whole farm by each turbine. It thus could be feasible even for a large farm.

## IV. Alleviating the Burden of Coordination

The cost of coordination per subsystem, which is quantified by Proposition 3.2, might happen to be steep. In such a case we may aim at trading off coordination and local performances. If the problem is localized, i.e., it concerns only a small number of subsystems, it can be resolved by tuning weights $\mu_i$ (cf. Remark 3.2). Yet if the problem is global, some alterations to either (7c) or the local criteria (3) are to be made. A possible approach would be to adjust $\bar{F}$ or/and $F_\alpha$ to render them closer to each other. This direction, however, is normally empirical and problem oriented and thus is not pursued here. Instead, in this section we assume that these gains are predetermined and consider some options of alleviating the burden of coordination on $\Sigma_i$'s via adjusting the global problem.

*A. Soft constraint formulation*

Coordination requirements may be taken into account via soft constraints. Namely, the minimization of (7a) under constraint (7c) may be substituted with

$$\text{minimize} \quad \mathcal{J} = \sum_{i=1}^{\nu} \mathcal{J}_i + \frac{\lambda}{1-\lambda} \|\bar{u} - \bar{F}\bar{x}\|_2^2 \quad (24a)$$

$$\text{subject to} \quad \Sigma_i, \quad i = 1, \ldots, \nu \quad (24b)$$

for some $\lambda \in [0, 1]$ and no constraints imposed on the behavior of the center of mass. The case $\lambda = 1$ effectively corresponds to the hard constraint formulation. Picking $\lambda < 1$ would mean that the coordination requirement is displaced with a coordination incentive. A satisfactory trade-off between local objectives and coordination can then be reached via tuning $\lambda$.

The arguments of §III-B apply to (24) mutatis mutandis[3], splitting the minimization of the coupled $\mathcal{J}$ into $\nu$ uncoupled problems. As in the hard constraint case, $\nu - 1$ of them are unconstrained LQR problems in terms of $\tilde{x}_i$ and $\tilde{u}_i$, $i = 2, \ldots, \nu$. The remaining problem, the one formulated in terms of $\tilde{x}_1 = \bar{x}$ and $\tilde{u}_1 = \bar{u}$, is now the LQR problem for (4) and the performance index

$$\int_0^\infty \begin{bmatrix} \bar{x}' & \bar{u}' \end{bmatrix} \begin{bmatrix} Q_\alpha + \frac{\lambda}{1-\lambda} \bar{F}'\bar{F} & -\frac{\lambda}{1-\lambda} \bar{F}' \\ -\frac{\lambda}{1-\lambda} \bar{F} & \frac{1}{1-\lambda} I_m \end{bmatrix} \begin{bmatrix} \bar{x} \\ \bar{u} \end{bmatrix} dt.$$

The resulting control law for the center of mass is:

$$\bar{u}(t) = (\lambda \bar{F} - (1-\lambda) B' X_\lambda) \bar{x}(t), \qquad (25)$$

where $X_\lambda \geq 0$ is the stabilizing solution of the ARE

$$(A + \lambda B \bar{F})' X_\lambda + X_\lambda (A + \lambda B \bar{F}) \\ + Q_\alpha + \lambda \bar{F}' \bar{F} - (1-\lambda) X_\lambda B B' X_\lambda = 0. \quad (26)$$

The overall controller is then in the same "diagonal plus rank-one" form (16), modulo replacing $\bar{F}$ with $\lambda \bar{F} - (1-\lambda) B' X_\lambda$.

*Remark 4.1:* The soft constraint formulation could, in principle, be viewed as a special case of (7). Indeed, the solution of Theorem 3.1 is recovered via the mere substitution $\bar{F} \to \lambda \bar{F} - (1-\lambda) B' X_\lambda$. Thus, formulation (24) brings no extra freedom to the design. Rather, we view it as a convenient means to trade off local and global goals. Moreover, the soft constraint formulation prompts extensions that are not covered by (7). One such extension will be considered in §IV-B. ▽

The following proposition quantifies the trade-off between coordination and the local performance for the $i$th subsystem.

*Proposition 4.1:* The stabilizing solution $X_\lambda$ of (26) satisfies $X_\alpha \leq X_\lambda \leq \bar{X}$ and $Y_\lambda := \frac{d}{d\lambda} X_\lambda \geq 0$. Furthermore, the optimal solution of (24) renders

$$\mathcal{J}_i = \mathcal{J}_{i,\text{opt}} + \mu_i^2 \bar{x}_0' (X_\lambda - \lambda(1-\lambda) Y_\lambda - X_\alpha) \bar{x}_0,$$

which never exceeds the quantity in Proposition 3.2, and

$$\sigma(\lambda) := \|\bar{u} - \bar{F}\bar{x}\|_2^2 = (1-\lambda)^2 \bar{x}_0' Y_\lambda \bar{x}_0.$$

*Proof:* It can be shown, by differentiating (26) and rearranging terms, that $Y_\lambda$ satisfies the Lyapunov equation

$$A_\lambda' Y_\lambda + Y_\lambda A_\lambda + (\bar{F} + B'X_\lambda)'(\bar{F} + B'X_\lambda) = 0, \qquad (27)$$

where $A_\lambda := A + B(\lambda \bar{F} - (1-\lambda) B' X_\lambda)$ is Hurwitz. This proves that $Y_\lambda \geq 0$. The first claim of the proposition then follows by the facts that $X_\alpha = X_\lambda|_{\lambda=0}$ and $\bar{X} = X_\lambda|_{\lambda=1}$.

The expression for $\mathcal{J}_i$ results from Proposition 3.2 by replacing $\bar{F} \to \lambda \bar{F} - (1-\lambda) B' X_\lambda$ (cf. Remark 4.1) and using the fact that under this choice $\bar{X} \to X_\lambda - \lambda(1-\lambda) Y_\lambda$ (can be verified by straightforward, albeit lengthy, algebra).

Finally, the control law (25) violates constraint (7c) by

$$\bar{u}(t) - \bar{F}\bar{x}(t) = -(1-\lambda)(\bar{F} + B'X_\lambda)\bar{x}(t) \\ = -(1-\lambda)(\bar{F} + B'X_\lambda)e^{A_\lambda t}\bar{x}_0.$$

[3]Theorem 4.2 in §IV-B presents a formal proof of a more general problem.

The expression for the norm of the constraint violation then follows by (27). □

Comparing the expressions for $\mathcal{J}_i$ given in Propositions 3.2 and 4.1, we can see that by relaxing the coordination constraint we reduce the cost of coordination for the $i$th subsystem by

$$\alpha_i(\lambda) := \mu_i^2 \bar{x}_0' (\bar{X} - X_\lambda + \lambda(1-\lambda) Y_\lambda) \bar{x}_0 \geq 0.$$

In fact, it can be shown that $\alpha_i(\lambda) = 0$ iff the cost of coordination in the original formulation $\mu_i^2 \bar{x}_0' (\bar{X} - X_\alpha) \bar{x}_0 = 0$ as well. In other words, whenever the coordination constraint (7c) does not come for free, formulation (24) alleviates its burden. Furthermore, it is readily seen that

$$\dot{\alpha}_i(\lambda) = \mu_i^2 \lambda \bar{x}_0' Z_\lambda \bar{x}_0 \quad \text{and} \quad \dot{\sigma}(\lambda) = (1-\lambda) \bar{x}_0' Z_\lambda \bar{x}_0,$$

where $Z_\lambda := (1-\lambda) \frac{d}{d\lambda} Y_\lambda - 2Y_\lambda \leq 0$ verifies

$$A_\lambda' Z_\lambda + Z_\lambda A_\lambda - 2(\bar{F} + B'X_\lambda - (1-\lambda) B' Y_\lambda)' \\ \times (\bar{F} + B'X_\lambda - (1-\lambda) B' Y_\lambda) = 0$$

and is uniformly bounded as a function of $\lambda$. Hence, we have that $\lim_{\lambda \to 1} \dot{\sigma}(\lambda) = 0$, whereas, in general, $\lim_{\lambda \to 1} \dot{\alpha}_i(\lambda) \neq 0$.

Thus, we may expect that a relatively small deviation from the ideal behavior of the center of mass may result in a relatively large reduction in the cost of coordination for the subsystems. As a matter of fact, at the other end of the range, at $\lambda = 0$, the picture is mirrored. Thus, by adding a slight coordination penalty to the global cost function $\sum_i \mathcal{J}_i$ we can introduce coordination with little effect on local performances.

### B. Frequency weighted soft constraints

In many situations, we might not be interested in coordinating the center of mass over all possible situation in local subsystems. For example, we can persuade coordination only in a low frequency range. This may be useful in applications where the required group behavior (e.g., power production of a wind power plant discussed in Section II) is slower than that of individual subsystems (e.g., dynamics of a wind turbine). Such situations can be accommodated by replacing the second term in the right-hand side of (24a) with the $L^2$ norm of the signal $z_\sigma$, satisfying

$$\begin{cases} \dot{x}_\phi(t) = A_\phi x_\phi(t) + B_\phi(\bar{u}(t) - \bar{F}\bar{x}(t)), & x_\phi(0) = 0 \\ z_\sigma(t) = C_\phi x_\phi(t) + D_\phi(\bar{u}(t) - \bar{F}\bar{x}(t)). \end{cases}$$

Thus, $z_\sigma$ is the signal $\bar{u} - \bar{F}\bar{x}$ filtered by

$$W_\phi(s) = D_\phi + C_\phi(sI - A_\phi)^{-1} B_\phi.$$

(without loss of generality we may assume that the realization of $W_\phi$ is minimal). This leads to the following problem:

$$\text{minimize} \quad \mathcal{J} := \sum_{i=1}^{\nu} \mathcal{J}_i + \|z_\sigma\|_2^2 \qquad (28a)$$

$$\text{subject to} \quad \Sigma_i, \quad i = 1, \ldots, \nu \qquad (28b)$$

The weighing filter $W_\phi$ aims at shaping the coordination penalty over different frequencies. Moreover, by choosing $W_\phi(s)$ with pure imaginary poles we can enforce hard coordination constraints at some frequencies.

To formulate the solution to (28) we need the ARE

$$A'_\sigma X_\sigma + X_\sigma A_\sigma + C'_\sigma C_\sigma - (X_\sigma B_\sigma + C'_\sigma D_\sigma) \\ \times (D'_\sigma D_\sigma)^{-1}(B'_\sigma X_\sigma + D'_\sigma C_\sigma) = 0, \quad (29)$$

where

$$\left[\begin{array}{c|c} A_\sigma & B_\sigma \\ \hline C_\sigma & D_\sigma \end{array}\right] := \left[\begin{array}{cc|c} A_\phi & -B_\phi \bar{F} & B_\phi \\ 0 & A & B \\ \hline C_\phi & -D_\phi \bar{F} & D_\phi \\ 0 & Q_\alpha^{1/2} & 0 \\ 0 & 0 & I \end{array}\right],$$

and the associated feedback gain

$$F_\sigma = \begin{bmatrix} F_{\sigma 1} & F_{\sigma 2} \end{bmatrix} := -(D'_\sigma D_\sigma)^{-1}(B'_\sigma X_\sigma + D'_\sigma C_\sigma),$$

partitioned compatibly. The following theorem is the main result of this subsection:

*Theorem 4.2:* Let $A + B\bar{F}$ be Hurwitz and $(Q_\alpha, A)$ have no unobservable pure imaginary modes. Then (29) has a stabilizing solution $X_\sigma \geq 0$ such that its $(2,2)$ block, partitioned compatibly with the partition of $A_\sigma$, satisfies $\bar{X} \geq X_{\sigma 22} \geq X_\alpha$, and the control law solving (28) is

$$u_i(t) = F_\alpha x_i(t) + \mu_i(\bar{F} - F_\alpha)\bar{x}(t) + \mu_i \bar{u}_\phi(t), \quad (30)$$

where $\bar{u}_\phi := M_\phi(F_{\sigma 2} - \bar{F})\bar{x}$ and

$$M_\phi(s) := I + F_{\sigma 1}(sI - A_\phi - B_\phi F_{\sigma 1})^{-1} B_\phi.$$

*Proof:* To shorten the exposition, we assume through the proof that $D_\phi = 0$, the general case follows by similar steps.

Following the arguments of §III-B, we rewrite the problem in terms of the aggregate variables $x$ and $u$. The dynamics of the aggregate system are now coupled,

$$\begin{bmatrix} \dot{x}_\phi \\ \dot{x} \end{bmatrix} = \begin{bmatrix} A_\phi & \mu' \otimes (-B_\phi \bar{F}) \\ 0 & I_\nu \otimes A \end{bmatrix} \begin{bmatrix} x_\phi \\ x \end{bmatrix} + \begin{bmatrix} \mu' \otimes B_\phi \\ I_\nu \otimes B \end{bmatrix} u,$$

and the cost function is uncoupled (not if $D_\phi \neq 0$):

$$\mathcal{J} = \int_0^\infty \left( x'_\phi C'_\phi C_\phi x_\phi + x'(I_n \otimes Q_\alpha)x + u'u \right) dt.$$

The dynamics of the plant can still be decoupled via transformation (11) with $U$ satisfying (12). It is readily verifiable that the transformed dynamics are now

$$\begin{bmatrix} \dot{x}_\phi \\ \dot{\tilde{x}} \end{bmatrix} = \begin{bmatrix} A_\phi & e'_1 \otimes (-B_\phi \bar{F}) \\ 0 & I_\nu \otimes A \end{bmatrix} \begin{bmatrix} x_\phi \\ \tilde{x} \end{bmatrix} + \begin{bmatrix} e'_1 \otimes B_\phi \\ I_\nu \otimes B \end{bmatrix} \tilde{u}$$

and the weights matrices of the criterion remain unchanged. Thus, we again end up with $\nu$ separate problems. The last $\nu - 1$ of them are exactly the same problems in terms of $\tilde{x}_i$ and $\tilde{u}_i$ for $i = 2, \ldots, \nu$ as in the case studied in Section III. The first one is the LQR problem for the plant

$$\dot{x}_\sigma = A_\sigma x_\sigma + B_\sigma \tilde{u}_1, \quad \text{where } x_\sigma := \begin{bmatrix} x_\phi \\ \tilde{x}_1 \end{bmatrix}$$

and the cost function $\tilde{\mathcal{J}}_1 = \|C_\sigma x_\sigma + D_\sigma \tilde{u}_1\|_2^2$. This problem is well defined. Indeed, *(i)* the pair $(A_\sigma, B_\sigma)$ is stabilizable by the controllability of $(A_\phi, B_\phi)$ and the first assumption of the theorem $(A_\sigma + B_\sigma \begin{bmatrix} F_\phi & \bar{F} \end{bmatrix}$ is Hurwitz iff $A_\phi + B_\phi F_\phi$ is Hurwitz); *(ii)* the observability of $(C_\sigma, A_\sigma)$ and the second assumption guarantee that the realization $D_\sigma + C_\sigma(sI - A_\sigma)^{-1} B_\sigma$ has no imaginary axis invariant zeros. The optimal solution of the LQR above is then the static state feedback

$$\tilde{u}_1 = F_\sigma x_\sigma = F_{\sigma 1} x_\phi + F_{\sigma 2} \tilde{x}_1,$$

where $F_\sigma$ is generated by the stabilizing solution of (29). Because $\tilde{x}_1 = \bar{x}$, $\tilde{u}_1 = \bar{u}$, and $x_\phi = (sI - A_\phi)^{-1} B_\phi(\bar{u} - \bar{F}\bar{x})$, the state feedback above can be expressed as follows:

$$\tilde{u}_1 = \left( \bar{F} + M_\phi(s)(F_{\sigma 2} - \bar{F}) \right) \tilde{x}_1.$$

The control law (30) in the original coordinates follows then by repeating the steps of the proof of Theorem 3.1. □

Comparing (30) and (16), we can see that the effect of replacing the hard constraint (7c) with filtered soft constraints amounts to adding the signal $\bar{u}_\phi$ to the control law. Because the zeros of $M_\phi(s)$ are exactly the poles of $W_\phi(s)$, the spectrum of $\bar{u}_\phi$ vanishes at the frequencies where the weight goes to infinity (the imaginary poles of $W_\phi(s)$), recovering the hard constraint case. This shows that we can indeed enforce hard coordination constraints at certain frequencies by using weighs with $j\omega$ poles.

The expressions for the coordination mismatch and the local costs of coordination are not as transparent as those analyzed in the previous subsection. We thus just present the formulae for their computation. To that end, note that, as follows from the proof of Theorem 4.2, the state vector of the center of mass under the control law (30) is the impulse response of a system with the transfer function

$$G_{\bar{x}}(s) = E'_2(sI - A_\sigma - B_\sigma F_\sigma)^{-1} E_2 \bar{x}_0,$$

where $E_2 := \begin{bmatrix} 0 \\ I \end{bmatrix}$. Hence, $\bar{u}_\phi$ is the impulse response of

$$M_\phi(s) G_{\bar{x}}(s) = (F_\sigma - \bar{F} E'_2)(sI - A_\sigma - B_\sigma F_\sigma)^{-1} E_2 \bar{x}_0.$$

The coordination mismatch is $\bar{u} - \bar{F}\bar{x} = \bar{u}_\phi$, so its energy

$$\|\bar{u} - \bar{F}\bar{x}\|_2^2 = \bar{x}'_0 X_{\phi 22} \bar{x}_0,$$

where $X_{\phi 22}$ is the $(2,2)$-subblock of the solution of the Lyapunov equation

$$(A_\sigma + B_\sigma F_\sigma)' X_\phi + X_\phi (A_\sigma + B_\sigma F_\sigma) \\ + (F_\sigma - \bar{F} E'_2)'(F_\sigma - \bar{F} E'_2) = 0.$$

Now, the control law (30) is of the form $F_\alpha x_i + v_i$ for $v_i = \mu_i((\bar{F} - F_\alpha)\bar{x} + \bar{u}_\phi)$. Following the arguments used in the proof of Proposition 3.2, the cost of coordination for the $i$th subsystem is $\|v_i\|_2^2$. This quantity is the energy of the impulse response of

$$\mu_i(\bar{F} - F_\alpha) G_{\bar{x}}(s) + \mu_i M_\phi(s) G_{\bar{x}}(s) \\ = \mu_i(F_\sigma - F_\alpha E'_2)(sI - A_\sigma - B_\sigma F_\sigma)^{-1} E_2 \bar{x}_0.$$

Thus,

$$\mathcal{J}_i = \mathcal{J}_{i,\text{opt}} + \mu_i^2 \bar{x}'_0 X_{\nu 22} \bar{x}_0,$$

where $X_{\nu 22}$ is the $(2,2)$-subblock of the solution of

$$(A_\sigma + B_\sigma F_\sigma)' X_\nu + X_\nu (A_\sigma + B_\sigma F_\sigma) \\ + (F_\sigma - F_\alpha E'_2)'(F_\sigma - F_\alpha E'_2) = 0.$$

It can be shown, e.g., via the use of the return difference equality, that $X_{\nu 22} \leq \bar{X} - X_\alpha$, which is to be expected. Details of this derivation, however, are beyond the scope of this paper.

## C. Wind farm example (cont'd)

From a load balancing perspective, it is the slow variations in power mismatch that are troublesome, whereas the fast ones are considered relatively benign. To account for this, we consider an alternative formulation, where the power tracking requirement is relaxed at high frequencies. This is achieved via the formulation (28) with $B = B_u$, $Q_\alpha = C'_z C_z$, $\bar{F} = 0$, $\mu_i = 1/\sqrt{\nu}$, and

$$W_\phi(s) = \frac{\sqrt{\lambda/(1-\lambda)}}{s} \quad (31)$$

for $\lambda \in [0, 1]$. The integrator in $W_\phi$ guarantees zero net DC-power deviation. Indeed, by Theorem 4.2 the solution is

$$u_i = F_\alpha x_i - F_\alpha x_\mathrm{a} + u_\mathrm{a},$$

where $x_\mathrm{a}$ is the average state (as in §III-D), $u_\mathrm{a} = M_\phi F_{\sigma 2} x_\mathrm{a}$, and the filter

$$M_\phi(s) = \frac{s}{s + \omega_\sigma}$$

where $\omega_\sigma = -F_{\sigma 1}$. The high-pass form of $M_\phi$ ensures that the spectrum of $u_\mathrm{a}$ vanishes at the zero frequency. Since $u_\mathrm{a}$ corresponds to the contribution of an average WT to the net power deviation, zero net power deviation is enforced at DC.

The solid line in Fig. 3 shows the trade-off between the cost of coordination per an individual turbine and $\|u_\mathrm{a}\|_2^2$. The

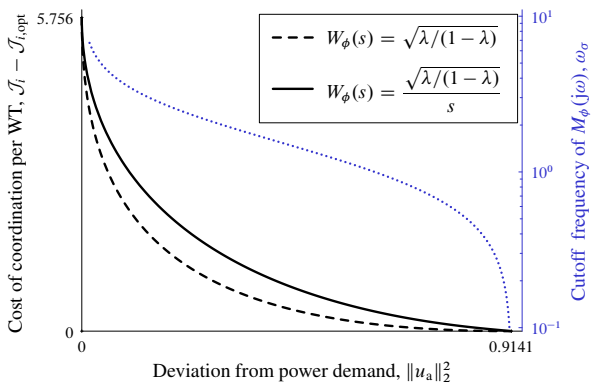

Fig. 3. Trade-off curves for local and coordination performances with $W_\phi = \sqrt{\lambda/(1-\lambda)}$ (dashed line) and $W_\phi = \sqrt{\lambda/(1-\lambda)}/s$ (solid line). The dotted blue line shows the cutoff frequency $\omega_\sigma$ of the filter $M_\phi$.

results show that a small relaxation of the power tracking requirement results in a relatively large improvement in individual WT performance. For comparison, we also present the trade-off curve for the formulation considered in §IV-A (the dashed line). This formulation corresponds to the static $W_\phi = \sqrt{\lambda/(1-\lambda)}$ in (28). We can see that the use of the static $W_\phi$ yields better coordination performance $\|u_\mathrm{a}\|_2$ for every level of deterioration of the local performances. This, however, may be expected, because the weight (31) effectively imposes hard constraints at the zero frequency for every $\lambda \neq 0$.

The dotted blue curve in Fig. 3 presents the cutoff frequency $\omega_\sigma$ of $M_\phi(j\omega)$. As the individual turbine performance improves, $\omega_\sigma$ decreases, which implies that less of the slow variations are removed from the net power deviation.

## V. CONCLUDING REMARKS

In this paper we have studied a class of LQR problems, where autonomous agents with identical dynamics seek to reduce their own costs while coordinating their center of mass (average behavior). We have shown that the solution to these problems has two important *scalable* properties. First, the problem decomposes into two independent LQR problems: one for a single uncoordinated agent and one for the center of mass, whose dynamics has the same dimension as those of individual agents. Hence, the computational effort required to obtain the solution is independent of the number of agents (since all agents are assumed to be identical, only one unconstrained LQR problems needs to be solved). Second, the structure of the resulting controller is transparent, comprising a (block) diagonal decentralized part and a (block) rank-one coordination term. The coordination term relies on information about all subsystems, but only requires a simple averaging operation. This renders the structure well suited for implementation in distributed control applications.

We have also revealed several other properties of the optimal solution. In particular, the cost of coordination incurred by each subsystem has been quantified and shown to vanish as the number of subsystems grows; the coordination problem has been interpreted in terms of a consensus-like cost function; the cost of the cumulative deviation of subsystems from the center of mass has shown to be independent of the behavior of the center of mass itself. We have also considered imposing coordination via soft constraints and quantified the trade-off between local and coordination performances in this case.

Although we have studied only the specific LQR problem, the diagonal-plus-low-rank structure may show up in a wider spectrum of applications. Relatively straightforward extensions include problems with $r$ coordination constraints (would result in a diagonal-plus-rank-$r$ configuration) and output-feedback $H^2$ formulations (adding local estimators). Other directions may be less trivial. For instance, it may be important to account for additional constraints on the information exchange between agents, like delays or a sampled-data structure. Another possible direction that might require a substantial alternation of the solution procedure is to consider coordination among heterogeneous agents. Furthermore, it is interesting to investigate the possibility of reducing information processing/complexity by imposing the diagonal-plus-low-rank structure in problems, where it does not arise as an outcome of the unconstrained optimization procedue.

Last but not least, up to this point we managed to discuss distributed control without mentioning the word "graph."

## ACKNOWLEDGMENTS

The authors owe a debt of gratitude to Bo Bernhardsson for posing thought-provoking questions and to Anders Rantzer for several helpful discussions.